\documentclass[11pt,letterpaper]{JHEP3}
\usepackage{epsfig}

\def\drawbox#1#2{\hrule height#2pt

        \hbox{\vrule width#2pt height#1pt \kern#1pt
              \vrule width#2pt}
              \hrule height#2pt}

\def\Fund#1#2{\vcenter{\vbox{\drawbox{#1}{#2}}}}
\def\Asym#1#2{\vcenter{\vbox{\drawbox{#1}{#2}
              \kern-#2pt       
              \drawbox{#1}{#2}}}}

\def\funda{\Fund{6.5}{0.4}}

\def\symm{\funda\kern-0.4pt\funda}

\newcommand{\half}{\frac{1}{2}}
\newcommand{\del}{\partial}

\newcommand{\be}{\begin{equation}}
\newcommand{\ee}{\end{equation}}
\newcommand{\bea}{\begin{eqnarray}}
\newcommand{\eea}{\end{eqnarray}}

\newcommand{\eq}[1]{equation~(\ref{#1})}
\newcommand{\eqs}[2]{equations~(\ref{#1}) and~(\ref{#2})}

\newcommand{\Lm}{{\mathcal L}}
\newcommand{\Sm}{{\mathcal S}}
\newcommand{\Hm}{{\mathcal H}}
\newcommand{\da}{\dot{a}}

\def\a{\alpha}
\def\as{\alpha_s}
\def\ag{\alpha_{GUT}}

\def\vp{\varphi}
\def\mpl{M_{P}}
\def\mg{M_{GUT}}
\def\ms{M_{s}}

\def\gs{g_{s}}
\def\sp{\;\;\;,\;\;\;}

\def\k{\kappa}

\def\p{\phi}
\def\VEV#1{\left\langle #1\right\rangle}
\newcommand{\GeV}{\mbox{GeV}}
\def\Montreal{Montr\'eal}
\def\Quebec{Qu\'ebec}

\preprint{{\tt hep-th/0301138}\\ McGill 02-36 \\ SU-GP-03/10-1}

\title{Brane inflation and reheating}

\author{John H. Brodie\\
Perimeter Institute\\
35 King Street North\\
Waterloo, Ontario N2J 2W9, Canada\\
\email{jbrodie@perimeterinstitute.ca}}
\author{Damien A. Easson \thanks{Address prior to September 1, 2003:
Department of Physics,
McGill University,
\Montreal, \Quebec,  H3A 2T8, Canada}\\
Department of Physics \\
Syracuse University \\
201 Physics Building \\
Syracuse, NY 13244-1130, USA\\
\email{easson@physics.syr.edu}}

\abstract{We study inflation and reheating in a brane world model
derived from Type IIA string theory.  This particular setup is
based on a model of string mediated supersymmetry breaking.
The inflaton is one of the transverse scalars of a D4-brane which
has one of its spatial dimensions
stretched between two
NS5-branes, so that it is effectively three-dimensional.
This D4-brane is attracted to a D6-brane that is separated
from the 5-branes by a fixed amount. The potential
of the transverse scalar due to the D4/D6 interaction makes a good inflaton potential. As the
D4-brane slides along the two 5-branes towards the 6-brane it
begins to oscillate near the minimum of the potential.
The inflaton field couples to the massless
Standard Model fields through Yukawa couplings.
In the brane picture these couplings are introduced by having
another D6'-brane
 intersect the D4-brane such that the 4-6' strings, whose lowest lying
 modes are the Standard Model matter, couple to
the scalar mode of the 4-4 strings, the inflaton.
The inflaton can decay into scalar and spinor particles on
 the 4-6' strings, reheating
the universe. Observational data is used
to place constraints on the parameters of the model.}

\keywords{D-branes, Cosmology of Theories beyond the SM}

\begin{document}

\baselineskip16pt
\parskip=4pt

\section{Introduction}\label{intro}
Recently there has been significant progress made in the field of
string cosmology (see e.g. \cite{Easson:2000mj}, \cite{Quevedo:2002xw}).
Arguably, among the greatest triumphs are the stringy realizations
of the inflationary universe paradigm.~\footnote{For references to such
models see \cite{Quevedo:2002xw},\cite{Jones:2002cv} and the references therein.}
In this paper we construct an inflationary brane world scenario from Type IIA
string theory.~\footnote{Other inflationary models based on IIA string theory
are \cite{Herdeiro:2001zb} and \cite{Kyae:2001mk}.}
The novel feature of this model is the reheating mechanism.
Our brane world inflates during its motion through a bulk spacetime.  In this way,
the setup shares some of the features of ``Mirage Cosmology"
\cite{Kehagias:1999vr}.~\footnote{Inflation in Mirage Cosmology was studied
in \cite{Papantonopoulos:2000yz,Papantonopoulos:2000xs}.}
The motion is caused by the exchange of massless closed strings \cite{Dvali:1998pa} (gravitons) between
the D4-brane, whose transverse scalar plays the role of the inflaton,
and a bulk D6-brane.  Inflation ends when the  D4-brane
begins to rapidly oscillate in the bulk about the minimum of the inflaton potential.
The inflaton field couples via Yukawa interactions to the massless
Standard Model (SM) fields that live on strings stretched between the inflaton
D4-brane and another D6'-brane.  The inflaton also couples gravitationally to
KK modes and massive string states. We find that the
reheat temperature is too low to excite massive closed string states.
Oscillations of the D4-brane excite
the 4-6' strings of which the lightest modes are the Standard Model fields.
The inflaton $\phi$ decays into scalar $\chi$ and
spinor fields $\psi$ on the 4-6' strings during the reheat process.
The interaction Lagrangian
giving rise to these decays appears naturally in string theories.
The inflationary phase is analyzed using the
standard slow-roll approximations (see, e.g.~\cite{Lyth:1998xn}).
Observational data is used to place constraints on the parameters of the model.
Finally, we briefly comment on the nature of the dark matter in this scenario.

\section{Brane dynamics}\label{dynamics}
Our concrete starting point involves the branes of ten-dimensional
Type IIA string theory. In \cite{Brodie:2001pw}, a similar configuration was considered
within the context of ``string
mediated" supersymmetry breaking in a Hanany-Witten model with rotated D6-branes.
This model allows us to break supersymmetry at the string scale  in some very heavy non-MSSM
messenger fields (the 4-6 strings) and then communicate the SUSY breaking to the
inflaton field living on the D4- brane world at a lower scale. Although the original motivation
for string mediation was to communicate supersymmetry breaking to the Standard Model fields,
here we use it to generate a potential for the inflaton field as well.

Consider a D4-brane
along the directions 1236, stretched in the 6-direction between two NS5-branes
extended in 12345.  We take the distance between the NS5-branes, $L_6$,
to be finite.  Separated from the NS5-branes in the 7-direction
is a D6-brane extended in 123689. The directions 456789 are
compact so that we have $3+1$ dimensional
gravity coupled to the massless fields on the brane (i.e. the D4-brane
appears 4-dimensional as long as we consider energies smaller
than $L_6^{-1}$).  This setup is depicted in Fig. 1.~\footnote{Note
that in the complete setup we include orientifold planes, which will be discussed
below and are not included in the figure.}

\EPSFIGURE[r]{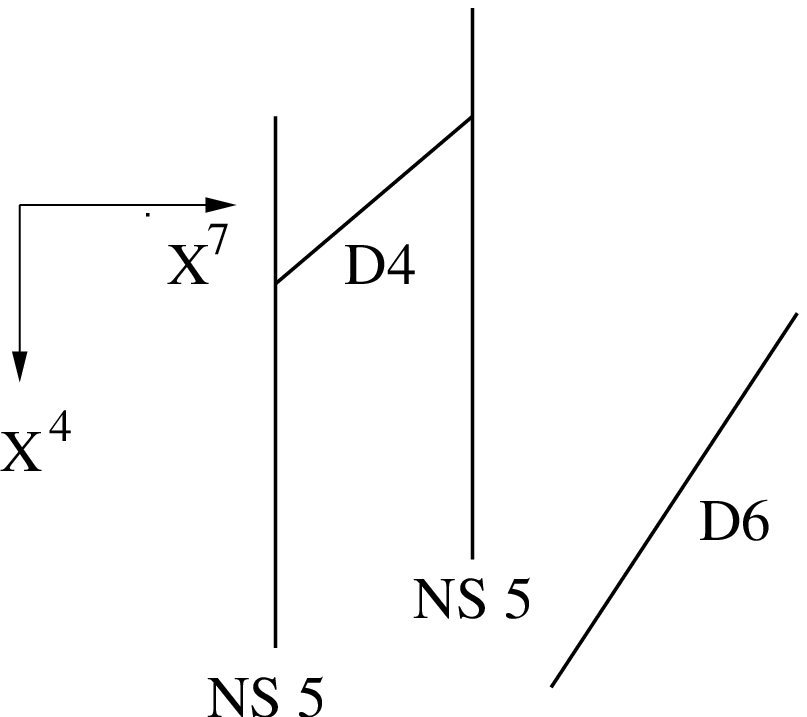,width=2in,height=2in}{Depicted is a D4-brane
in the 1236 directions, suspended between two NS5-branes extended in 12345.  The
D4-brane is gravitationally attracted towards a D6-brane extended in 123689. The D4-brane and the
NS5-branes are BPS, and so the D4-brane does not feel their background.}\label{fig1}

Here we discuss the details of this brane configuration.
The background metric and dilaton for the D6-brane are given by
\begin{eqnarray}
ds^2 = f^{-1/2}dx^2_{||} + f^{1/2}dy^2\\
e^{-2\Phi} = f^{3/2} \\
f(r) = 1+{g_s\sqrt\alpha'\over 2r} \label{harmonic}
\,,
\end{eqnarray}
where $x_{||}$ are the coordinates parallel to the brane and $y$ are the coordinates
perpendicular to the brane.
The fluctuations of the D4-brane are described by the Dirac-Born-Infeld action
\be\label{dbi}
\Sm =-\int{e^{-\Phi}T_4d^5x \sqrt{{\rm -det} (G_{ab}+B_{ab}+2\pi\alpha' F_{ab})}}
\,,
\ee
where $T_4={M_s^5\over (2\pi)^4g_s}$ is the tension of the
D4-brane.  Here $G_{ab}$ and $B_{ab}$ are the pull-back
of the spacetime metric and anti-symmetric 2-form field to the brane
 and $F_{ab}$ is the gauge field
living on the brane~\cite{Polchinski:1996na}.
The background values of the antisymmetric 2-form and
the gauge field on the D4-brane are taken to be zero ($F=B=0$). The metric $G$
of the D4-brane is induced by the metric of the
D6-brane
\be
G_{ab} = h_{ab} + d_aX^id_bX^jg_{ij} = f^{-1/2}\eta_{ab} + f^{1/2}d_aX^id_bX_j
\,,
\ee
where the $X^i$ are the
coordinates transverse to the brane.  The indices take on the values $i=4,5,7,8,9$ and
$a = 0, 1, 2, 3$. The determinant of the metric can be written
\begin{eqnarray}
{\rm det} G_{ab} = {\rm det} (h_{ac}\eta_{cb}) {\rm det} (\eta +
h^{ab}g_{ij}d_aX^id_bX^j) \,.
\end{eqnarray}
For small fluctuations in the $X$ coordinates, we can write
\begin{eqnarray}
{\rm det} G = {\rm det} (h){\rm det}(\eta ) ( {\rm det} \eta + Tr h^{-1}g dXdX + \cdots) \nonumber \\
= f^{-{5\over 2}}(-1 + f dXdX + \cdots)
\end{eqnarray}
The Lagrangian for the fluctuations of the dimensionally reduced D4-brane in the background of a D6-brane is,
therefore,
\be\label{sbrane}
\Sm = \int d^4x\sqrt{H} T_4L_6(-f^{-1/2} + {1\over 2} f^{1/2}H^{-1}d_aX^4 d^aX^4 + \cdots)
\,,
\ee
where $H$ is the (flat) metric seen by the low energy fields on the brane and
we have dimensionally reduced the D4-brane in the $x^6$ direction by placing the
D4-brane in the interval ($L_6$) between two NS 5-branes.  In the above, we focus on
the $X^4$ coordinate along the NS5-brane because
$\p\equiv \ms^2 X^4$ acts like an inflaton on the D4-brane.
Let us define, $S \equiv M_s^2r$, where $r$ is the distance between the 6-brane and
the 4-brane.  We take the D4-brane to lie along the NS5-branes separated a
distance $A\alpha'$ from the D6-brane where $A>>M_s$. $A$ here is
non-dynamical in the limit we are considering. Notice that $S=\sqrt{\phi^2 + A^2}$
never reaches zero for any value of $\phi$. Hence, the 4-6 string can never be shorter than
the string scale and will not become tachyonic.
Using these definitions and substituting the expression for the harmonic function $f$ (\ref{harmonic}),
into the action (\ref{sbrane}) gives the Lagrangian of the brane in terms of $\p$,
\be\label{lbr}
\Lm_{brane} = T_4L_6\left(\half \ms^{-4} \sqrt{1+\frac{\gs\ms}{2\sqrt{\p^2 + A^2}}} (\del \p)^2
- \frac{1}{\sqrt{1+\frac{\gs\ms}{2\sqrt{\p^2 + A^2}}}} \right)
\,.
\ee

\section{Brane cosmology}\label{bcos}
In order to discuss cosmology on the brane we need the complete action on the brane.
The full Type IIA bulk Lagrangian is \begin{eqnarray}
\label{massiveIIA}
 \Sm &=& {1\over 2\kappa_{10}^2} \int d^{10}x\,
  \sqrt{-g}\left\{e^{-2\Phi}\Big[R + 4|d\Phi|^2
  -{1\over 2}|H|^2\Big] - {1\over 2}| {G}_2|^2
  - {1\over 2}|\widetilde{G}_4|^2\right\} \nonumber\\[5pt]
 & + & {1\over 4\kappa_{10}^2} \int \left\{
    \mbox{}{1\over 2}G_4G_4B-{1\over 2}G_2G_4B^2 + {1\over 6}G_2^2B^3
    \right\} ,
\end{eqnarray}
where the gauge invariant field strengths are given by
\begin{eqnarray}
 H &=& dB, \nonumber \\
  {G}_2 &=& dC_1, \\
 \widetilde{G}_4 &=& dC_3 - C_1\wedge H  \;. \nonumber
\end{eqnarray}
In the above, $\Phi$ is the dilaton, $B$ is NS-NS antisymmetric tensor 2-form
potential, $C_1$ and $C_3$ are the Ramond-Ramond 1-form and 3-form potential
respectively.
Dimensionally reducing to four-dimensions on a torus of radius $R$
and focusing on the most relevant massless modes (the graviton and
the dilaton), the bulk action becomes
\be \Sm = {1\over
2\kappa_{4}^2} \int d^{4}x\,
  \sqrt{-g}\left\{e^{-2\Phi} R + 4|d\Phi|^2
 \right\}
    \ee
where \be {1\over 2\kappa_{4}^2} = {R^6\over 2\kappa_{10}^2} \ee
A conformal transformation of the metric takes us from the string frame to
the Einstein frame.
\be g_{\mu\nu}^s =  g_{\mu\nu}^E
e^{\Phi\over 2} \ee
In this frame Newton's constant is independent
of time and position.
The bulk action becomes
\be \Sm_E = {1\over
2\kappa_{4}^2} \int d^{4}x\,
  \sqrt{-g}\left\{ R + 2|d\Phi|^2
 \right\}
 \,.
 \ee
The full action becomes
\be\label{ltot} \Sm = \int d^4x \sqrt{-g} \big(-\frac{R}{2\k_4^2} +
\Lm_{D4-brane}+ \Lm_{D6-brane}+ \Lm_{NS5-brane}+ \Lm_{O6-brane}+
\Lm_{O5-brane}\big) \,, \ee where the first term is just the
Einstein-Hilbert action and the second term is given by \eq{lbr}.
We comment on the other contributions to the action below.
Strictly speaking the D4-brane action given by
\eq{lbr} is in the string frame which is incompatable with the
Einstein-Hilbert action. However, we will always be working in
a regime where the dilaton is approximately constant, that is to say the
large $r$ limit of dilaton profile in equation (2.2). In this limit there
is only an order 1 difference between the D4-brane action in the
 string frame and the Einstein frame.~\footnote{The D4-brane in Einstein frame is
 $ S=\int d^4x \sqrt{H_E} T_4L_6 (-f^{-19/8}+{1\over 2}f^{-5/8}H_E^{-1}dXdX+...)$.}

The D6-brane and the NS5-branes have scalar
fields on their world volumes that can act as ``inflaton" fields.
However, these branes are heavier than the D4-branes and therefore
do not move significantly on the time scales considered here. The three-dimensional
energy density of the D6-brane compared to that of the D4-brane can be calculated using
numerical values we find at the end of this paper. We find
\be {\rho_{D6}\over \rho_{D4}}={(M_s^7/g_s)R_6R_8R_9\over (M_s^5/g_s)L_6}=
 10^2 \ee Since acceleration is two
time derivatives of position we expect that the time scales on
which the branes move will go like the square root of the mass
density. \be {t_{D6}\over t_{D4}} \simeq \sqrt { {\rho_{D6}\over
\rho_{D4}}} = 10 \ee
The D6-brane and the NS5-branes also carry tension
which couples to gravity and can contribute to inflation. In order
to eliminate their effects we have added orientifold planes which
have negative tensions. Orientifolds do not fluctuate and so they do not introduce any new
dynamical fields.

For the physics we are interested in, the only dynamically relevant
scalar field in the brane configuration is  the motion of the D4-brane.
Note that in the above
action we have made a critical assumption:
the moduli of the compact space and the dilaton are stabilized by
some unknown physics.~\footnote{After the completion of this paper we became aware of
some recent progress made in solving the problem of moduli stabilization
\cite{Giddings:2001yu, Kachru:2003aw}.}  We will comment further on this
point below.

We take $\p$ to be spatially homogeneous and time dependent and
the metric on the brane is the Friedmann-Robertson-Walker (FRW) metric
\be
ds^2 = n(t)^2 dt^2 - a(t)^2 \left[ \frac{dr^2}{\sqrt{1-kr^2}} + r^2 d\Omega^2 \right]
\,,
\ee
where $n$ is the lapse function, $a$ is the scale-factor on the brane and $d\Omega^2$ is
the metric on the unit sphere.  For simplicity, we will assume a flat universe with
$k=0$.~\footnote{Note that during inflation the effects of non-zero $k$ will be very
quickly inflated away because the terms containing $k$ in the Friedmann equation
are diluted by $a^{-2}(t)$.}

The equations of motion on the brane are determined by varying the action (\ref{ltot})
with respect to $n$, $a$ and $\p$. Varying with respect to the lapse function (which we set to unity in
the EOM) gives
\be\label{infit}
H^2 = \frac{\sqrt{2}\k_4^2 T_4 L_6 \left(2(A^2 + \p^2)(2\ms^4 + \dot \p^2) + \gs \ms\sqrt{A^2 + \p^2}
\dot \p^2 \right)}{12 \ms^4 (A^2 + \p^2) \sqrt{2 + \frac{\gs\ms}{\sqrt{A^2 + \p^2}}}}
\,,
\ee
where $H \equiv \da(t)/a(t)$ is Hubble parameter.
It is easy to see from \eq{infit} that, in the slow-roll regime
(large $\p$ and small $\dot \p$), inflation
will occur on the brane
\be
a(t) = \exp{\Hm t} \,,
\ee
where $\Hm = \k_4 \sqrt{T_4 L_6/3}$.

\subsection{Slow-roll inflation}
The standard approximation technique for analyzing inflationary models is the
slow-roll approximation.
We assume that the inflaton $\p$
is initially large.  At this point, $\p$ is on a relatively flat part of its potential,
and therefore, $\dot \p$ is small.  A plot of the entire potential $V(\p)$ is given
in Fig. 2.
In the slow-roll regime $\p$ satisfies
$\gs\ms/2S << 1$, ($\p \simeq S$), and \eq{lbr} becomes
\be
\Lm_{brane} \approx T_4 L_6 \left(\half \ms^{-4} (\del \p)^2 - (1-\frac{\gs\ms}{4\p}) \right)
\,.
\ee
By rescaling $\p$ so that $\ms^2 \vp = \sqrt{T_4 L_6}  \p$, and defining
$M \equiv \gs (T_4 L_6)^{3/2}/4\ms$, the full Lagrangian on the brane becomes
\be
\Lm_{tot} \simeq \sqrt{-g} \left( -\frac{R}{2 \k_4^2}
    + \half (\del_a \vp)^2 - V(\vp) \right)
\,,
\ee
where
\be\label{pot}
V(\vp) = T_4 L_6 - \frac{M}{\vp}
\,.
\ee
>From the potential (\ref{pot}), we can calculate the
slow-roll parameters
\bea
\varepsilon \equiv \half \mpl^2 (V'/V)^2 \simeq \half (\frac{\mpl M}{ L_6 T_4 \vp^2})^2 \\
\eta \equiv \mpl^2 (V''/V) \simeq \frac{-2M \mpl^2}{ L_6 T_4 \vp^3} \,,
\eea
where $\mpl$ is the four-dimensional Planck mass and the prime denotes differentiation
with respect to $\vp$.  In order for the slow-roll approximation to be valid, the
inflaton must be on a region of $V$ which satisfies the flatness conditions
$\varepsilon <<1$ and $|\eta|<<1$.  Inflation ends when $\vp = \vp_{end}$ (when $|\eta|$ becomes
of order one),
\be
\vp_{end}^3= \left(\frac{2 \mpl^2 M}{L_6 T_4}\right)
\,.
\ee
The amount of inflation is given by the ratio of the scale factor at the final time $t_{end}$ to
its value at some initial time $t_i$.  The total number of e-folds is
\bea
N \equiv \ln{\frac{a(t_{end})}{a(t_i)}} = \int_t^{t_{end}} H dt \simeq
\int_{\vp_{end}}^{\vp} \frac{V}{V'} d\vp \nonumber \\
\simeq \frac{2}{3}\Big(\frac{\vp}{\vp_{end}}\Big)^3 = \frac{L_6 T_4}{3 M \mpl^2} \vp^3
\,.
\eea
We take the desired number of e-foldings to be $N=58$.  This implies
the value of $\vp$ when interesting scales cross outside the horizon is
\be
\vp \simeq 4.43 \, \vp_{end}
\,.
\ee
The power spectrum measured by COBE at the scale $k\simeq 7.5 H_0$ is~\cite{Lyth:1998xn}
\be
\delta_H \equiv \frac{2}{5} \mathcal{P}_R^{1/2} = 1.91 \times 10^{-5}
\,.
\ee
In terms of the potential $V$, we have
\be
\delta_H^2(k) = \frac{1}{75 \pi^2 \mpl^6} \frac{V^3}{V'^2}
\,,
\ee
where $V$ and $V'$ are evaluated at the epoch of horizon exit for the scale $k = a H$.
Using this expression it is possible to place constraints on the parameters of the potential,
\be\label{constraint}
\mpl^{-3} \frac{V^{3/2}}{V'} = 5.3 \times 10^{-4}
\,.
\ee
Using (\ref{constraint}) we find
\be\label{julia}
L_6 \propto \left(\frac{\gs}{N}\right)^2 \left(\frac{\mpl}{\ms}\right)^5 \frac{1}{\ms}
\,.
\ee

The spectral index, defined by $n -1 = 2\eta -6 \varepsilon$, is shifted slightly to the red
(for $N> 4/3$),
$n \simeq 1 - 4/3N$.  For $N=58$ the spectral index is $n \simeq .98$. The variation of $n$ with respect to
wave number $k$ is
\be
\frac{dn}{d\,\ln{k}} = 2 \xi^2 = 2 \mpl^4 \frac{V' V''}{V^2}\simeq \frac{4}{3N^2} = .0004
\,.
\ee

\subsection{Cosmological constant}
When $\p=0$ the brane world reaches the bottom of its potential
and we may approximate the Lagrange density (\ref{lbr}) as
\bea\label{small} \Lm_{brane} \simeq T_4L_6\left(\half \ms^{-4}
(c_1 - c_2 \p^2) (\del \p)^2 - (\frac{1}{c_1} + c_3 \p^2 )\right) \,,\\
c_1=\sqrt{1+\frac{\gs\ms}{2A}}=\frac{\gs\ms}{8A^3}c_2^{-1} =
(\frac{\gs\ms}{8A^3}c_3^{-1})^{1/3} \,.
\eea
Here we are confronted
with a non-standard kinetic term, $\Lm_{kin} \propto \p^2 (\del
\p)^2$. To place the Lagrangian into canonical form, we define \be
\del \Psi = \ms^{-2} \sqrt{T_4 L_6 c_1 (1 - \frac{c_2}{c_1} \p^2)}
\; \del \p \,. \ee In terms of the new inflaton variable $\Psi$,
\eq{small} is
\be\label{small2} \Lm_{brane} \simeq \half (\del\Psi)^2 - m^2
\Psi^2 -\Lambda \,,
\ee
where $m^2 = \ms^2 \sqrt{\frac{T_4
L_6}{c_1}}c_3$. Note the presence of a non-zero cosmological
constant $\Lambda = T_4 L_6/c_1$.  We do not attempt to address
the cosmological constant problem in this paper. One way to avoid this problem
is to fine-tune away $\Lambda$ by adding a negative
cosmological constant to \eq{ltot}. This might be achieved in
string theory by embedding our brane set-up in $AdS_4\times
S^7$.  A sketch of the entire inflaton potential is obtained by considering
equations (\ref{pot}) and (\ref{small2}) (see Fig. 2).

\EPSFIGURE[r]{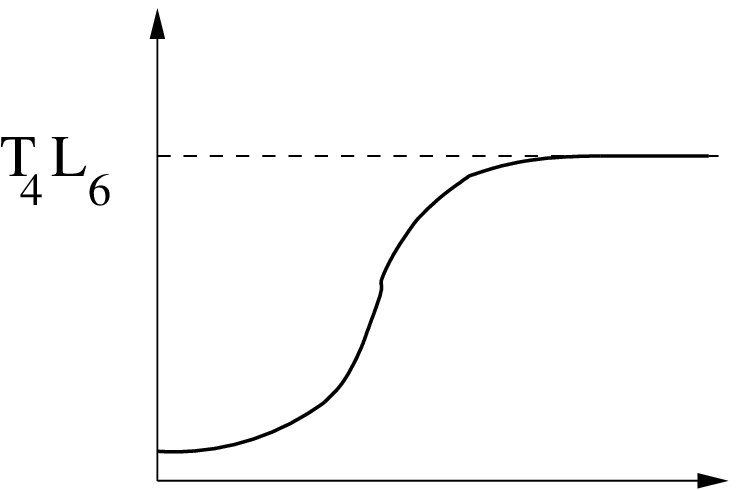,height=1.5in}{This is a sketch of the inflaton
potential $V(\phi)$.}\label{fig2}

\section{The string scale}
Empirical evidence (and assuming the MSSM is correct) seems to indicate that the
gauge couplings of the Standard Model unify at the GUT scale
$\mg \simeq 2 \times 10^{16} \, \GeV$, $\ag \simeq 1/25$~\cite{Dimopoulos:1981yj}.
After compactification to four-dimensions, the Planck mass
$\mpl = (8 \pi G_N)^{-1/2}=2.42 \times 10^{18} \GeV$,
is determined by dimensional reduction (see e.g., \cite{Shiu:1998pa})
\be\label{volume}
g_s^2 \mpl^2 = \frac{\ms^8 V_{6d}}{(2\pi)^6 \pi}
\,,
\ee
where $V_{6d}$ is the 6d compactification volume. The one compact dimension of the D4-brane we live on
has volume $2\pi R_6$.  The factor of $(2\pi)^6 \pi$ in the above comes from $G_{10}$
in type I string theory \cite{Sakai:1987jg},
\be
8\pi G_{10} = \frac{g_s^2 (2\pi)^7}{2 \ms^8}
\,.
\ee
The string coupling $\gs=e^{\VEV{\Phi}}$ is related to the gauge coupling by
\be\label{vol}
\gs = 2 \ms (2 \pi L_6) (2\pi)^{-1}\as
\,.
\ee
Using \eqs{constraint}{vol} it is possible to determine the relation between $M_s$ and $\delta_H$.
At the perturbative level, string theory does not stabilize the dilaton.  It is therefore likely that
such stabilization will arise from non-perturbative effects (i.e when $\gs \gtrsim 1$).
Taking $\gs=1$, $\a \simeq 1/25$ and $N=58$, we find
\be
\ms \simeq 6.5 \times 10^{15} \,\GeV
\,,
\ee
which is within the allowed range, between $1 \,\GeV$ and $10^{16}\,\GeV$
\cite{Arkani-Hamed:1998rs,Antoniadis:1998ig}  . (Note that for
$N\lesssim 20$, it is possible to have $\ms \simeq \mg \simeq 10^{16}\,\GeV$.)
Using this value of $M_s$ and \eq{vol}, we find $L_6 = (25/2) \ms^{-1}$.

The branes have a thickness of order the string scale.
So that the branes have enough room to move freely in the extra dimensions we take
$R_5$, $R_7$, $R_8$ and $R_9$ to be large (i.e., $R_i>M_s^{-1}$, where $i=5,7,8,9$).
For simplicity we assume that all $R_i \simeq R$.  We assume that the remaining compact
dimensions ($R_4$ and $R_6$) are compactified with
volume $(2\pi/\ms)$ (i.e. at the self-dual value) and $2 \pi L_6$ respectively. Note
that if the radius is less than the self-dual value ($R_4< M_s^{-1}$),
then the T-dual description is more appropriate~\cite{Jones:2002cv}.
In the T-dual scenario,
\be
g_s \rightarrow g_s R\a^{-1} \ms^{-1} \sp R_4 \rightarrow R_4^{-1} \ms^{-2}
\,.
\ee
Here the T-dual torus has $R_4 \geq \ms^{-1}$.  Therefore, it is possible
to consider the cases where $R_4 \geq \ms^{-1}$, with $R_4 = \ms^{-1}$.

The entire six-dimensional compact space has volume
\be
V_{6d} = (2 \pi \ms^{-1})(2\pi L_6) (2\pi)^4 (R_5 R_7 R_8 R_9)
\,.
\ee
>From \eq{volume} we find
\be
R_i^4 \simeq \left(\frac{\mpl}{\ms}\right)^2 \frac{2 \pi}{25 \ms^4}
\,,
\ee
from which
$R_5^{-1}, R_7^{-1}, R_8^{-1}, R_9^{-1} \simeq 4.8 \times 10^{14}\, \GeV$.
Note that in general, as long as all of the compactified dimensions are larger than the string
scale by at least
a factor of two, cosmologically dangerous defects will be too heavy to
be produced~\cite{Jones:2002cv}.

\section{Reheating the brane}
There are now several different realizations of brane inflation 
\cite{Herdeiro:2001zb,Kyae:2001mk,Dvali:1998pa,Jones:2002cv,Dasgupta:2002ew,inf:lation}.
The details of reheating in these models has not been fully explored, although
in general inflation ends due to tachyon condensation (either by brane-antibrane annihilation
or brane-brane collisions).~\footnote{See the 
models \cite{Herdeiro:2001zb,Dasgupta:2002ew} for two exceptions.} 
The process of tachyon condensation may lead to severe cosmological
problems~\cite{Kofman:2002rh}. Our model does not suffer from these difficulties
since the D4 and D6 branes never get close enough for the 4-6 string to become tachyonic.
Instead, the reheating period in this model begins when $\phi$ rolls to zero and
oscillates about the minimum of its potential.  Physically, this
corresponds to the brane world bouncing up and down around the
point $\phi=0$ on the NS5 brane.  In realistic models, chiral
fields make up part of the brane modes and will couple to the
gauge fields.  During the reheat process the inflaton will decay
into these chiral fields. Since, in this picture, the inflaton is
a brane mode the reheating process should be quite efficient.
~\footnote{For detailed discussions of the reheating process, see
e.g.
\cite{Traschen:1990sw,Kofman:1994rk,Shtanov:1994ce,Kofman:1997yn}.}
Consider the inflaton scalar $\p$ interacting with a scalar field
$\chi$ and a spinor field $\psi$ that live on the brane by the
interaction \be\label{interact} \Lm_{int} = -\half g^2 \p^2 \chi^2
- h \bar\psi\psi\p \,. \ee The total decay rate of $\p$ particles
into the scalar $\chi$ and spinor $\psi$ particles is \be \Gamma =
\Gamma(\p \rightarrow \chi\chi) + \Gamma(\p \rightarrow \psi\psi)
\,. \ee Following the arguments of \cite{Kofman:1997yn}, the decay
products of the inflaton are ultrarelativistic and their energy
density decreases due to the expansion of the univers much faster
than the energy of the oscillating field $\p$.  Therefore, the
reheating phase ends when the Hubble constant $H\simeq 2/3t$,
becomes smaller than $\Gamma$.  This corresponds to a universe of
age $t_{RH}\simeq \frac{2}{3}\Gamma^{-1}$.  The age of the
universe with energy density $\rho$ is $t=\sqrt{\mpl/6\pi\rho}$.
Hence, the energy density at the end of reheating is \be
\rho(t_{RH}) \simeq \frac{3\mpl^2}{8\pi}\Gamma^2 \,. \ee Assuming
thermodynamic equilibrium sets in quickly after the decay of $\p$,
the matter on the brane reaches the ``reheat temperature" defined
by \be\label{rh} T_{RH} \simeq \left(\frac{30 \rho(t_{RH})}{\pi^2
n} \right)^{1/4} \simeq 0.2 \sqrt{\Gamma \mpl} \,, \ee where $n$,
the number of relativistic degrees of freedom, is expected to be
$n(T_{RH}) \sim 10^2 - 10^3$.  In our model, there are two ways in
which the inflation can decay: One is through gravitational,
closed string modes which we'll call $\Gamma_G$ and the other is
via the Yukawa couplings which constitute open string modes which
we'll call $\Gamma_h$. $\Gamma_G$ can be estimated using a formula
in \cite{Kyae:2001mk} \be \Gamma_G = {G_N^2M\over M_s^6V_{6d}} \ee
Plugging our numbers into this formula and using it to calculate
the reheat temperature we find $T_{RH} \simeq 10^8$GeV. There is
also the decay channel via the Yukawa coupling $\Gamma_h$. We
estimate this to be \be \Gamma_h \simeq hM^{1/5}\ee Since $h$ is a
tunable parameter in this model we can make this as small as we
like. Namely, if we choose $h=10^{-14}$, we can make $\Gamma_h$ an
order of magnitude bigger than $\Gamma_G$ such that it is the
preferred decay channel while still keeping the reheat temperature
below $10^{9}$GeV. In this way we avoid overproduction of
gravitinos during reheating. ~\footnote{In this analysis we have
neglected the possibility of parametric resonance effects which
can enhance the rate of boson particle production, preheating
\cite{Kallosh:1999jj}. We thank R.~Brandenberger for pointing out
this possibility.} Because the reheat temperature is not very
large compared to $\mpl$, the creation of undesirable defects
seems unlikely.

Equation (\ref{interact}) can appear naturally in our string model in the
following way: Introduce a D6'-brane along directions 123(45)7(89) where the
parenthesis means that the brane is at an angle $\theta_{48}$ in the 48 direction
and an angle $\theta_{59}$ in the 59 direction. As long as the angle with the
original D6-brane in the 123689 satisfies $\theta_{48}+\theta_{59} +\theta_{67} = 0$
then the D6-branes will be BPS \cite{Berkooz:1996km}. We can choose the position of
the D6'-brane such that it intersects the D4-brane at $X^4=0$
and the 4-6' strings are massless. The 4-6' strings in this scenario will play
the role of the Standard Model matter with the Standard Model gauge fields living on the D4-branes. When
the D4-brane begins to oscillate about its minimum, it will excite the
4-6' strings that couple to it. If however, $\theta_{48}={\pi/2}$, that is to say the D6' is perpendicular to
the D6-brane, then there is no
part of the D6'-brane transverse to the D4-brane, and the D4-brane is not
coupled to the 4-6' strings. Thus the angle   $\theta_{48}$ is proportional to the
Yukawa coupling between the 4-4, 4-6', and the 6'-4 strings.
 In superspace notation there is a coupling of the
form \be W = h \int d^2\theta \Phi Q\tilde Q \ee  where $\theta_{\alpha}$ is
the superspace coordinate, $\Phi = \phi + \theta \psi_{\phi} + \theta^2F_{\phi}$,
is part of the supermultiplet living on the 4-4 string and $Q = q + \theta\psi + \theta^2F_q $
the supermultiplet living on the 4-6'. The F-terms are not dynamical and can
be integrated out. The potential in components
reduces to \be V = h\phi^2q^*q  + h\phi\psi\bar\psi \ee
where $h = \sqrt 2 {\rm cos}{\theta_{48}}$.
In this model the Standard Model fields will naturally get a mass of order
the supersymmetry breaking scale which is the same as the inflaton mass. In such a situation
energy conservation would forbid the inflaton from decaying into Standard Model fields via
the Yukawa interaction proposed. However, one can fine tune the Standard Model mass to be light
to avoid this problem. In the brane model, the mass of the Standard Model fields are generated by the
bending of the D4-brane by the D6-brane. This has the effect of stretching the
4-6' strings giving the fields on them a mass proportional to the stretched distance.
The fine tuning amounts to moving the D6'-branes in the $x^7$ direction such that
the 4-6' strings are again light.

\EPSFIGURE[r]{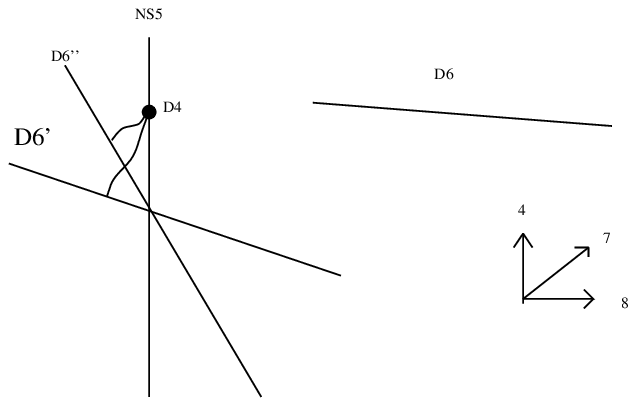}{Here we see a D4-brane couples to a D6'-brane and a D6''-brane.
The D4-brane slides back and forth along the NS5-brane, and as
it passes by the D6' and D6''-brane, it excites the 4-6' and 4-6'' strings.
These represent the SM fields and dark matter respectively. }

This suggests a way to have dark matter fields: Introduce another D6''-brane in the same direction
123(45)7(89) displaced from the D6'-brane in the $x^6$ direction.
Now the D4-brane also couples to the 4-6'' fields and so the inflaton field can decay into
these fields. We can write an interaction
\be V = t\phi^2p*p + t\phi\omega\bar\omega + h\phi^2q^*q  + h\phi\psi\bar\psi \ee
where $p$ is a scalar field, $\omega$ is a fermion, and
$t$ is the Yukawa coupling determined be the angle $\theta_{48}$ between
the D6''-brane and the D6-brane. We can have $t/h$ very large by tuning the
angles appropriately. In this way we can get more dark matter then visible matter.
However, this dark matter will not be dark because it couples to the Standard Model gauge fields
which we have taken to live on the D4-branes. To solve this problem we could put the Standard Model gauge
fields on the D6'-branes, but this would modify our cosmological analysis. We leave the
details of this model of dark matter for future work. The brane configuration for this
model is shown in Fig. 3.

The inflaton can also decay into massive Kaluza-Klein modes gravitationally.
These KK modes are heavy of a mass $10^{14}$ GeV and decay rapidly into light Standard Model fields.
Massive closed string states are slightly heavier than the reheat temperature and
so are not produced.
In Fig. 4, we see the gravitational interaction of the inflaton field
with KK modes.

\section{Conclusions}

In this paper we have studied a stringy realization of the
inflationary universe model.  The scenario is formulated within
the context of Type IIA string theory and the separation between
D-branes plays the role of the inflaton.  Similar models were
explored previously (see e.g., \cite{Quevedo:2002xw,Jones:2002cv}
for an extensive list of references). The standard slow-roll
approximation technique was used to study the inflationary
dynamics. Using observational (COBE) data, we placed constraints
on the parameters of the model.  For a desirable 58 e-foldings of
inflation (and taking $\gs \simeq 1$) the spectral index is
shifted to the red (by an acceptable amount) $n \simeq .98$.  The
string scale is found to be $\ms \simeq 10^{15}\GeV$.  The ``size"
of the D4 brane in the $x_6$-direction (distance between the NS5
branes) is $L_6^{-1} \simeq 5.2 \times 10^{14}\,\GeV$. The scales
of the six compactified dimensions are $R_4^{-1}\simeq
10^{15}\,\GeV$, $R_6^{-1} = L_6^{-1}$ and $R_5^{-1}, R_7^{-1},
R_8^{-1}, R_9^{-1} \simeq 4.8 \times 10^{14}\, \GeV$. We propose a
novel mechanism for reheating within the brane world context.
Reheating occurs when our brane world rapidly oscillates about the
minimum of its potential. Massive string states can be created
during the reheating period as well as KK modes in the $T^6$ which
can then decay into the massless Standard Model fields that live
on the brane.  The reheat temperature is estimated to be $T_{RH}
\lesssim 10^{8} \GeV$ avoiding overproduction of gravitinos during
reheating . Compared to $\mpl$, the reheat temperature is probably
not large enough to create dangerous defects which would overclose
the universe. In our model the cosmological constant problem
remains and it must be fine-tunned away in the usual fashion.

\EPSFIGURE[r]{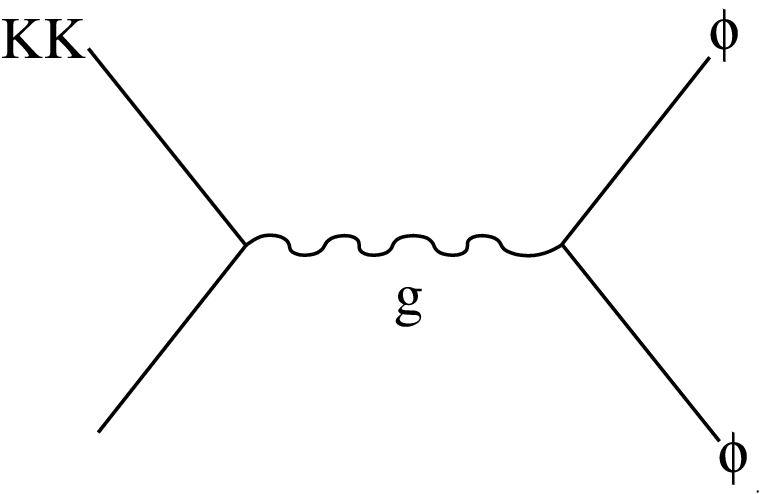,height=1.5in}{The Feynmann diagram of an inflation field
interacting with  KK modes via a graviton.}\label{fig4}

Finally, we comment on the possible nature of dark matter in this model.
It is possible to introduce an additional D6''-brane (the dark matter brane)
that couples to the inflaton D4-brane through a Yukawa interaction in the same way as the
inflaton couples to the Standard Model fields. By tuning the relative angle of the D6''-brane with
the NS5-brane we can change the Yukawa couplings. In this way we can easily arrange to have
the inflaton field decay into the appropriate ratio of dark matter to visible matter. By displacing
the D6''-brane slightly from the D4-brane we can give the 4-6'' strings a small mass.
This mass can be fine tuned to be much smaller than the string or supersymmetry breaking
scale. To get the dark matter to be ``dark'', we had to consider moving the Standard Model gauge fields
from the D4-brane to the D6'-brane. If this was done and the gauge field on the
D4-brane was taken to be $U(1)$, the $U(1)$ gauge field on the
D6''-brane and the D4-brane will be free in the IR.
Thus the 4-6'' strings make for good dark matter because they are light massive particles
that do not interact with the Standard Model fields or themselves at low energies.

Other dark matter candidates are the 6-6 strings that live on the displaced D6-brane.
The inflaton field couples to these fields gravitationally. We can have $N_{D6}$
D6-branes giving us a $SU(N_{D6})$ gauge symmetry living on them. Supersymmetry breaking
would give a mass to the gauginos. The inflaton could then decay into these fermionic fields.
Since the D6-brane fields only couple gravitationally to the SM fields, they would be invisible
to observers on our brane. A Higgs field would have to be introduced to break the gauge
symmetry such that the gauginos do not interact strongly at low energies. The number of
D6-branes $N_{D6}$ could be adjusted such that one gets a large enough amount of dark matter.
We leave a more detailed discussion of this model to future work.

\section*{Acknowledgements}
We would like to thank R. Brandenberger, C. Burgess, B. Kyae, J. Maldacena,
A. Mazumdar and
G. Moore for helpful discussions.
DE is supported in part by NSF-PHY-0094122 and other funds from Syracuse
University. DE also acknowledges support from McGill University prior
to September 1, 2003.

\end{document}